Book Chapter

# Pedagogical Promise and Peril of AI: A Text Mining Analysis of ChatGPT Research Discussions in Programming Education

Juvy C. Grume[1*], John Paul P. Miranda[1], Aileen P. De Leon[1], Jordan L. Salenga[1], Hilene E. Hernandez[1], Mark Anthony A. Castro[1], Vernon Grace M. Maniago[1], Joel D. Canlas[1], Joel B. Quiambao[1]

1. **Pampanga State University**, Pampanga, Philippines

*** Correspondence:**
Juvy C. Grume, Pampanga State University, jncruz@pampangastateu.edu.ph

## ABSTRACT

GenAI systems such as ChatGPT are increasingly discussed in programming education, but the ways in which the research literature conceptualizes and frames their role remain unclear. This chapter applies text mining to publications indexed in a leading academic database to map scholarly discourse on ChatGPT in programming education. Term frequency analysis, phrase pattern extraction, and topic modeling reveal four dominant themes: pedagogical implementation, student-centered learning and engagement, AI infrastructure and human–AI collaboration, and assessment, prompting, and model evaluation. The literature prioritizes classroom practice and learner interaction, with comparatively limited attention to assessment design and institutional governance. Across studies, ChatGPT is positioned both as a learning aid that supports explanation, feedback, and efficiency and as a pedagogical risk linked to overreliance, unreliable outputs, and academic integrity concerns. These findings support responsible integration and highlight the need for stronger assessment and governance mechanisms.

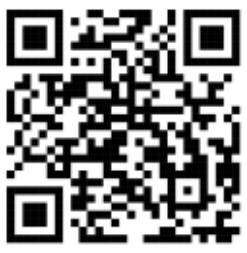

# INTRODUCTION

Programming remains one of the most demanding subjects in computer science education. Despite continued curriculum and pedagogical reforms, many students still struggle to learn programming effectively. Studies across multiple institutions report failure and withdrawal rates ranging from 30 to 50% in introductory programming courses (Margulieux et al., 2020; Simon et al., 2019). Students often describe programming as abstract and mentally challenging because it requires logical reasoning, syntax accuracy, and problem decomposition (Dirzyte et al., 2023; Malik & Coldwell-Neilson, 2017). These challenges lower confidence and increase anxiety, which frequently result in disengagement and attrition (Dirzyte et al., 2023; Garcia, 2024). These persistent difficulties motivate educators to seek approaches that provide timely feedback, guided practice, and individualized support for novice programmers.

Earlier responses to these difficulties included intelligent tutoring systems and automated assessment platforms. These tools adapted feedback and evaluated student work through structured mechanisms. They improved performance in topics such as loops, recursion, and data structures (Crow et al., 2018; Kouam & William, 2024). However, they were limited in scale and unable to support complex or open-ended programming tasks. The steady growth of class sizes in computer science programs reduced the capacity of instructors to deliver individualized assistance. This situation created a need for scalable systems capable of providing interactive and immediate instructional support.

The emergence of large language models, particularly ChatGPT, redirected this pursuit. ChatGPT produces explanations, examples, and code corrections through natural conversation. Instructors now integrate it into lessons to help students understand logic and structure in programming, while learners use it to receive guidance, generate code, and verify solutions. Several studies report that ChatGPT supports motivation, coding accuracy, and self-efficacy when used with structured pedagogical frameworks (Garcia, 2025a; Wang & Fan, 2025). Other studies caution that unstructured use can lead to dependency, incorrect understanding, and ethical concerns regarding originality and fairness (Agbo et al., 2025; Muringa, 2025).

Within programming education, ChatGPT now functions as a tutor, debugging assistant, and feedback tool (Agbo et al., 2025; Gleeson, 2024). Students appreciate its accessibility and responsiveness, while researchers note issues related to plagiarism, misinformation, and unequal access to advanced versions (Reihanian et al., 2025; Tong, 2024). These perspectives demonstrate both optimism and caution, showing that ChatGPT combines pedagogical potential with ethical uncertainty.

This chapter examines this dual character through a text mining analysis of research on ChatGPT in programming education. It describes how recent publications discuss its educational applications and clarifies how ChatGPT can operate as both a productive learning aid and a source of academic and ethical challenges. The chapter aims to provide direction for teachers, researchers, and policymakers in creating responsible and effective strategies for integrating generative AI into programming instruction.

# BACKGROUND

Artificial intelligence has supported programming education for several decades through systems that provide adaptive feedback and automated assessment. Early tools such as intelligent tutoring systems and code visualization platforms improved learning performance but were limited in flexibility and coverage (Crow et al., 2018; Kouam & William, 2024). Automated graders improved efficiency by comparing program outputs with expected solutions, but they could not evaluate problem-solving logic or coding style (Garcia, 2025b). These earlier systems established the groundwork for AI-supported instruction but lacked the conversational and contextual capacity of current large language models (Agbo et al., 2025; Franklin et al., 2025).

Generative AI offers a more interactive and context-aware approach to programming instruction. ChatGPT responds to questions, explains code, and produces examples aligned with the learner's level of understanding. Current literature documents its use in tutoring, content development, and automated assessment (Franklin et al., 2025; Garcia, 2025b; Gleeson, 2024). Studies in programming education show that ChatGPT can generate functional code, clarify compiler messages, and simplify complex programming structures in different languages such as Python, Java, and C++ (Kosar et al., 2024; Phung et al., 2023; Shanto et al., 2025; Sun et al., 2024). Controlled applications report higher motivation, engagement, and self-perceived competence among students who use ChatGPT as a supplement to teacher-led instruction (Sun et al., 2024; Villegas-Ch et al., 2025). This body of evidence suggests that generative AI strengthens student participation and supports comprehension of difficult programming concepts.

Despite these advantages, concerns remain about ethical use and educational quality. Many students rely on AI-generated code without understanding its logic (Cotton et al., 2023; Silva et al., 2024). Teachers continue to face challenges in confirming authorship and ensuring that assessment practices remain fair and valid (Reihanian et al., 2025; Shafii & Berger, 2025; Swiecki et al., 2022). Some researchers note that excessive reliance on ChatGPT reduces persistence and limits independent debugging ability (Humble et al., 2024; Monib et al., 2024). These concerns demonstrate the need for structured guidelines and institutional policies that define appropriate AI use in academic settings.

The growing body of literature on ChatGPT remains limited in scope. Most studies focus on small classes or isolated instructional settings and seldom connect pedagogical and ethical dimensions within one analytical framework (Garcia, 2025; Kosar et al., 2024). Evidence about learning benefits and dependency patterns remains inconsistent, and institutional readiness for AI integration continues to vary (Agbo et al., 2025; Taheri et al., 2025). Few studies discuss ChatGPT's role as a long-term instructional partner or supervised learning resource in programming courses.

A comprehensive synthesis is needed to explain the dual role of ChatGPT in programming education. This chapter addresses that need through text mining of published studies to identify recurring educational and ethical themes. It provides a consolidated perspective on how researchers describe the advantages and risks of using ChatGPT in programming instruction. This background establishes the context for the text mining analysis that follows.

## MAIN FOCUS OF THE CHAPTER

This chapter focuses on how existing research describes the use of ChatGPT in programming education within its pedagogical, ethical, and institutional dimensions. It examines how scholars and educators conceptualize ChatGPT as a learning tool that supports practice, feedback, and problem-solving in programming courses. The discussion aims to clarify how current studies describe the balance between its educational benefits and the challenges that accompany its integration. The chapter also explores how ChatGPT is represented in the broader academic discourse on computer science education. It considers how researchers frame ChatGPT within discussions of instructional innovation, academic integrity, and student autonomy. The analysis seeks to understand how these perspectives influence teaching practices and institutional policies in programming education. Additionally, this chapter provides an overview of how the literature shapes the academic conversation about ChatGPT as both an opportunity and a challenge in programming education. It establishes the foundation for the text mining analysis, which examines the structure, themes, and key concepts that define current scholarly discussions on ChatGPT in computer science education.

## Method

The dataset consisted of 229 documents retrieved from Scopus using the search query TITLE-ABS-KEY(chatgpt AND (“*programming education” OR “teaching programming*” OR “*computer science education*”)). Only open-access journal articles and conference papers were included to ensure transparency and accessibility of research sources. After the removal of duplicate, irrelevant, and incomplete entries, 81 documents remained for initial review. A second validation excluded 12 documents that did not address programming education or the educational use of ChatGPT. The final dataset contained 69 documents that represented distinct academic discussions related on ChatGPT in programming and computer science education.

Each document underwent text preprocessing to prepare the corpus for analysis. All text was converted to lowercase, and punctuation marks, digits, and non-alphabetic symbols were removed. The corpus was tokenized to separate words, and standard English stopwords were removed to retain only semantically relevant terms. Porter stemming was applied to normalize word variations, such as reducing students and student to a single root form. These steps produced a clean and consistent dataset suitable for computational text analysis.

Three analytical procedures were applied. Term frequency analysis identified the most frequently occurring words to determine the dominant research concepts. Phrase pattern analysis through bigrams and trigrams examined co-occurring terms that represented conceptual relationships such as “*AI tools*,” “*programming education*,” and “*problem-solving skills*.” Latent Dirichlet Allocation (LDA) topic modeling was used to identify underlying themes in the dataset. The model was configured with four topics after multiple iterations for coherence and interpretability. Each document was assigned to a primary topic based on the highest probability score, and representative authors were recorded to link computational patterns with the corresponding scholarly discussions. Furthermore, to complement the computational process, each document was reviewed manually to identify reported opportunities, challenges, and limitations. This procedure ensured that the interpretation of results reflected both quantitative and contextual aspects of the academic discourse.

## Text Mining Analysis Results

### Term Frequency Analysis

The term frequency analysis showed a consistent focus on students, AI integration, and programming education (Table 1). The most frequent words were student, ChatGPT, AI, education, and program. These terms indicate that researchers describe ChatGPT as both a learning companion and a pedagogical tool. The presence of tool, learn, and use shows that ChatGPT is often discussed as an instructional support that encourages engagement and assists in understanding programming concepts. This pattern confirms that programming education remains the main context of scholarly attention and that AI integration is viewed as a significant component of teaching and learning in computer science.

**Table 1.**
*Top 10 stemmed terms.*

| Rank | Term | Frequency |
|---|---|---|
| 1 | student | 443 |
| 2 | chatgpt | 324 |
| 3 | ai | 291 |
| 4 | educ | 255 |
| 5 | program | 254 |

| Rank | Term | Frequency |
|---|---|---|
| 6 | use | 224 |
| 7 | learn | 222 |
| 8 | studi | 192 |
| 9 | tool | 149 |
| 10 | code | 143 |

## Phrase Pattern Analysis

The phrase pattern analysis identified key relationships among concepts within the corpus (Table 2). Phrases such as programming education, AI tools, and use ChatGPT appeared most frequently, showing that research often centers on formal classroom environments where AI supports coding practice and instruction. Phrases including problem solving, learning outcomes, and future research demonstrate that authors combine pedagogical interest with evaluation and reflection. The trigram ChatGPT programming education occurred most frequently, linking pedagogical themes with the technological basis of generative AI. The repetition of AI generated content and AI generated code indicates continuing attention to originality and plagiarism. These patterns reveal that current discussions are organized around three central areas: pedagogy, ethics, and learning outcomes.

**Table 2.**
*Top 10 bigrams and trigrams*.

| Rank | Bigram | Frequency | Trigram | Frequency |
|---|---|---|---|---|
| 1 | programming education | 59 | chatgpt programming education | 12 |
| 2 | ai tools | 38 | large language models | 9 |
| 3 | use chatgpt | 33 | ai generated content | 8 |
| 4 | computer science | 32 | computer science education | 8 |
| 5 | using chatgpt | 24 | tools like chatgpt | 6 |
| 6 | chatgpt programming | 21 | problem solving skills | 6 |
| 7 | ai generated | 20 | chatgpt computer programming | 6 |
| 8 | future research | 19 | benefits negative impacts | 5 |
| 9 | problem solving | 19 | potential benefits negative | 5 |
| 10 | learning outcomes | 18 | ai generated code | 5 |

## Topic Modeling Results

The LDA topic modeling analysis produced four themes that describe how research conceptualizes the use of ChatGPT in programming education (Table 3). These themes are *Pedagogical Use and Classroom Implementation, Student-Centered Learning and Engagement, AI Infrastructure and Human-AI Collaboration, and Assessment, Prompting, and Model Evaluation*. The first two topics represent the largest portion of the corpus and focus on the classroom and learner experience, while the remaining topics reflect growing attention to institutional and evaluative perspectives.

The topic *Pedagogical Use and Classroom Implementation* includes works by Husain (2024), Stoyanova et al. (2025), López-Fernández and Vergaz (2025), Yang et al. (2025), and Penney et al. (2025). These authors emphasize the importance of teacher facilitation, structured integration, and ethical awareness in classroom applications. Programming education research continues to concentrate on these learner-centered and instructional perspectives that frame ChatGPT as an aid to teaching and as a guide for practice. The topic *Student-Centered Learning and*

*Engagement* includes studies by Annuš (2025), Boudia and Krismadinata (2024), Raihan et al. (2025), Güner and Er (2025), and Troussas et al. (2025). These studies describe ChatGPT as an interactive tutor that supports motivation, engagement, and coding performance. These discussions by Annuš, Troussas et al., Güner and Er, and Raihan et al. describe ChatGPT as a formative support system that enhances problem-solving and student participation. While discussions by Husain, Stoyanova et al., and Penney et al. emphasize that structured pedagogy and active teacher supervision are necessary for effective use. These perspectives highlight that the educational potential of ChatGPT relies on instructional alignment and teacher involvement in the learning process.

The themes *AI Infrastructure and Human-AI Collaboration* and *Assessment, Prompting, and Model Evaluation* expand the discussion to institutional and technical aspects. Habiballa et al. (2025), Dunder et al. (2024), da Silva et al. (2024), Phung et al. (2024), and Aruleba et al. (2025) examine how transparency, readiness, and accountability influence the integration of ChatGPT into institutional systems. Grandel et al. (2025), Lee and Song (2024), Cowan et al. (2023), Lau and Guo (2023), and Leinonen et al. (2023) address assessment-related issues through studies on prompt design, feedback generation, and model accuracy. Parallel discussions from Habiballa et al., da Silva et al., and Aruleba et al. describe AI systems as educational resources that require teacher preparation and ethical oversight, while Grandel et al., Lee and Song, and Lau and Guo explain how human moderation enhances the reliability of AI-assisted evaluation.

**Table 3.**
*Top terms per topic and their distribution.*

| Topic | Label | Representative Terms | Count | Proportion |
|---|---|---|---|---|
| 0 | Pedagogical Use and Classroom Implementation | chatgpt, student, use, educ, learn, studi, ai, tool, implement, teach | 13 | 0.19 |
| 1 | Student-Centered Learning and Engagement | student, program, ai, chatgpt, learn, educ, use, studi, tool, task | 34 | 0.49 |
| 2 | AI Infrastructure and Human-AI Collaboration | ai, educ, student, llm, human, gener, model, code, program, task | 16 | 0.23 |
| 3 | Assessment, Prompting, and Model Evaluation | code, model, llm, perform, prompt, evalu, gpt, test, gener, feedback | 6 | 0.09 |

## Benefits and Opportunities of Using ChatGPT in Programming Education

### Enhanced Learning and Skill Development

ChatGPT strengthens programming proficiency when integrated through structured instructional frameworks. The R5E model presented by Abouelenein et al. (2025) improved student performance and critical thinking compared with unstructured use. PyChatAI developed by Alanazi et al. (2025) delivered real-time bilingual feedback that improved debugging and conceptual understanding among Python learners. The mobile learning system examined by Atta et al. (2025) enhanced motivation, self-efficacy, and coding accuracy through guided learning activities. The GPT-based code review system designed by Lee and Joe (2025) reduced academic dishonesty and increased precision in feedback, while Yilmaz and Yilmaz (2023) reported gains in computational thinking and programming confidence. These studies show that ChatGPT contributes to programming education when supported by pedagogical guidance and teacher facilitation.

### Personalized Learning and Accessibility

ChatGPT supports adaptive and personalized learning that responds to individual progress and skill level. The R5E model developed by Abouelenein et al. (2025) produced higher post-test gains in computational thinking and coding performance. The automated grading study by Bernik et al. (2025) showed that ChatGPT-4 achieved a 0.91 correlation with instructor ratings and generated consistent and fair feedback. PyChatAI in Alanazi et al. (2025) improved examination scores through bilingual and adaptive guidance. The fuzzy memory model introduced by Troussas et al. (2025) generated feedback based on learner progress and achieved high retrieval accuracy. Studies by Lee and Joe (2025) and Lee and Song (2024) demonstrated that ChatGPT provided clear and relatable explanations for abstract programming concepts. Research by Abdulla et al. (2024), Liao et al. (2024), and Barambones et al. (2024) showed that ChatGPT-based scaffolding enhanced problem-solving and expanded opportunities for practice in simulated programming environments. These studies indicate that ChatGPT supports accessibility and individualized learning through adaptive feedback and context-sensitive guidance.

### Instructor and Institutional Efficiency

The integration of ChatGPT improves instructional efficiency and reduces workload for teachers and institutions. Bernik et al. (2025) demonstrated that ChatGPT-4 graded programming submissions with a 0.91 correlation to instructor scores while completing evaluations faster than manual grading. Grandel et al. (2025) reported that the GreAIter system decreased grading time by over 75 percent without compromising accuracy. The GPT-based code review module developed by Lee and Joe (2025) improved error detection, supported large classroom management, and reduced repetitive teacher feedback. The SQL exam design study conducted by Aerts et al. (2024) showed that ChatGPT-3.5 generated coherent exam questions and reduced preparation time for instructors. These results demonstrate that ChatGPT assists in improving efficiency, consistency, and scalability in programming assessment and feedback.

## Challenges, Risks, and Limitations

### Academic Integrity and Misuse Risks

Concerns about academic integrity remain central to the discussion of ChatGPT in programming education. The survey conducted by Stoyanova et al. (2025) showed that teachers regarded ChatGPT as a major factor contributing to exam dishonesty, while students expressed more neutral perspectives. The experiment by Akçapınar and Sidan (2024) reported that students who received ChatGPT assistance obtained higher grades but often copied inaccurate outputs from the model. Shirazi et al. (2025) developed an outlier detection method to identify irregular grade patterns, but their approach could not confirm individual cases of cheating. Pan et al. (2024) evaluated AI-content detectors and observed poor accuracy in distinguishing AI-generated code from human-written submissions. These studies show that ChatGPT can both improve and distort performance results, reinforcing the need for institutional policies, clear ethical frameworks, and monitoring mechanisms that maintain fairness in programming education.

### Negative Cognitive and Educational Impacts

Cognitive dependency and reduced persistence represent continuing challenges in the use of ChatGPT for programming education. Stoyanova et al. (2025) reported that both teachers and students observed risks related to memorizing incorrect explanations and losing motivation to solve problems independently. Sun et al. (2024) noted that students who used ChatGPT showed higher satisfaction but no measurable improvement in learning outcomes. Mezzaro et al. (2024) found that students who relied heavily on ChatGPT produced fewer and less effective software tests. Gandhi and Muldner (2025) observed limited reasoning when students explained AI-generated code. Aruleba et al. (2025) and Azoulay-Schwartz et al. (2024) emphasized that frequent reliance on AI weakens problem-solving effort and deep understanding. These studies demonstrate that while ChatGPT supports comprehension, unmoderated use may restrict the development of independent reasoning and persistence in programming (Lepp & Kaimre, 2025).

### Technical Limitations and Unreliability

Technical limitations affect the accuracy, reliability, and consistency of ChatGPT's outputs in programming tasks. Annuš (2025) reported that students frequently encountered incomplete or incorrect code responses. Lepp and Kaimre (2025) found a negative correlation between frequent ChatGPT use and programming test performance. Husain (2024) noted that instructors experienced inconsistent or irrelevant responses when using ChatGPT for code generation. Troussas et al. (2025) observed that the fuzzy memory model maintained better contextual accuracy than standard ChatGPT but remained constrained during extended sessions. Mezzaro et al. (2024) and Frankford et al. (2024) identified repeated errors in code explanations and hallucinated responses. Cowan et al. (2023) found that only 30% of ChatGPT outputs were fully usable, while failed conversations occurred frequently. These studies emphasize that ChatGPT requires additional human supervision, algorithmic refinement, and contextual calibration to ensure dependable integration into programming education.

### Ethical and Social Concerns

Ethical and social concerns influence how ChatGPT is perceived and implemented in programming education. The survey conducted by Da Silva et al. (2024) showed that while most students valued ChatGPT for its instructional support, they also raised issues of plagiarism, privacy, and unequal access. Boudia and Krismadinata (2024) reported that both teachers and students expressed reservations about dependency, misinformation, and reduced creativity. Abdulla et al. (2024) observed that students favored institutional monitoring to ensure responsible use. Aruleba et al. (2025) documented uncertainty about authorship and the ethical use of AI-generated work. Their study also showed that ChatGPT assisted learners with disabilities, which raised questions about equity when access was limited. Teachers in Boudia and Krismadinata (2024) emphasized that AI could support learning but could not replace human creativity and judgment. These studies demonstrate the need for institutional frameworks, equitable access, and ethical awareness programs that guide responsible use of ChatGPT in computer science education (Stoyanova et al., 2025).

## SOLUTIONS AND RECOMMENDATIONS

The responsible use of ChatGPT in computer science education requires clear policies, structured pedagogy, and institutional support. Universities should establish course-level guidelines that define appropriate and inappropriate applications of ChatGPT in programming activities, assignments, and assessments. These policies must distinguish between guided learning and academic dishonesty. Instructors should communicate these standards at the start of each course and reinforce them through continuous monitoring and feedback. Verification procedures such as code walkthroughs or oral assessments can help confirm authorship and ensure academic integrity. These steps allow ChatGPT to serve as a learning resource that promotes understanding rather than a shortcut that undermines authentic programming practice.

Pedagogical integration should focus on active engagement and computational reasoning. Instructors can use ChatGPT to design structured exercises that require students to compare AI-generated code with their own, identify logic errors, and justify algorithmic choices. This approach encourages analysis and supports deeper comprehension of programming structures. ChatGPT can also assist in the creation of instructional materials, debugging sessions, and guided practice tasks. To maintain instructional quality, teachers must develop competence in AI literacy, prompt construction, and evaluation of AI-supported outputs. Regular training and collaboration among faculty can strengthen instructional consistency and ethical application in programming courses.

Equity and infrastructure development are essential for sustainable AI integration. Institutions should provide equal access to generative AI tools through campus-wide licenses or laboratory installations to ensure that all students can benefit from technological support. Collaboration with developers can improve model transparency, reduce response errors, and create logs that track student-AI interactions for monitoring and research. Universities should include accessibility features such as voice support and multilingual feedback to assist learners with different needs. Through policy clarity, structured pedagogy, and inclusive infrastructure, computer science education can adopt ChatGPT responsibly while preserving fairness, integrity, and authentic skill development.

# FUTURE RESEARCH DIRECTIONS

Future research should investigate the long-term influence of ChatGPT on learning behavior, programming competence, and self-regulated learning among students in computer science education. Most current studies are limited to short-term classroom experiments. Longitudinal research can assess how consistent exposure to ChatGPT affects problem-solving skills, code quality, and persistence over multiple courses. These studies can determine whether regular use of AI assistance strengthens or reduces students' independent reasoning and debugging abilities. Research in this area can also evaluate how ChatGPT influences the development of higher-order skills such as abstraction, design thinking, and algorithmic reasoning.

Further research should compare instructional frameworks that integrate ChatGPT with existing programming pedagogy. Experimental and quasi-experimental designs can assess which combinations of AI-assisted and teacher-led instruction produce the most balanced outcomes in learning and motivation. Studies can analyze course implementations in areas such as software development, algorithms, and database systems to identify effective approaches for using ChatGPT as a learning partner rather than a replacement for instructor guidance. Research on assessment models can explore how reflective documentation, oral defense, or version-controlled project submissions maintain authenticity while accommodating AI-supported work.

Cross-institutional and cross-cultural research can expand understanding of equity, readiness, and policy adoption in AI-assisted learning environments. Comparative studies can reveal how resource differences, teacher confidence, and institutional infrastructure shape the quality of ChatGPT integration. Partnerships between educators and developers can improve prompt accuracy, reduce bias, and enhance the interpretability of AI-generated explanations in programming tasks. Studies that combine learning analytics with behavioral observation can identify usage patterns and inform intervention strategies. These research directions can advance responsible and data-informed practices for integrating ChatGPT into computer science education.

# CONCLUSION

ChatGPT has become an important development in computer science education because of its capacity to generate, explain, and evaluate code in real time. Its use in programming courses presents opportunities for personalized instruction, immediate feedback, and improved engagement. When integrated through structured pedagogical frameworks, ChatGPT can strengthen computational thinking, self-efficacy, and problem-solving performance. However, unregulated use can create dependency and reduce independent reasoning. These contrasting effects emphasize the need for well-defined integration that upholds academic integrity and authentic learning.

The role of institutions and educators remains central in achieving responsible implementation. Clear policies, transparent assessment procedures, and teacher training ensure that ChatGPT enhances instruction rather than replaces the learning process. Courses should align AI use with active learning strategies that require justification, code evaluation, and reflection. Institutions should also provide equitable access to AI resources and technical support to prevent inequality among students. Through careful regulation and guided use, universities can balance technological assistance with the goal of developing independent and competent programmers.

In the broader context of computer science education, ChatGPT represents both innovation and responsibility. Its integration challenges traditional ideas about authorship, creativity, and expertise in programming. Continued collaboration among educators, researchers, and developers is essential to ensure that generative AI supports human learning objectives. With ethical governance, sound pedagogy, and institutional readiness, computer science education can adopt ChatGPT as a constructive partner that enhances understanding and promotes reflective and meaningful learning.

## KEY TERMS AND DEFINITIONS

**ChatGPT:** An AI system that generates code, explanations, and feedback through natural language interaction.

**Programming Education:** The structured teaching of coding, algorithmic reasoning, and problem-solving in computer science.

**GenAI:** An AI technology that creates new content such as text, code, or images based on learned data patterns.

**Pedagogical Integration:** The structured inclusion of ChatGPT in programming instruction to support learning and assessment.

**Student-Centered Learning:** An instructional approach that promotes active participation and individualized guidance through ChatGPT.

**Computational Thinking:** A problem-solving process involving abstraction, decomposition, and algorithmic reasoning.

**Active Learning:** A method that engages students in hands-on programming and evaluation of code solutions.

**Academic Integrity:** An ethical standard that ensures honesty and originality in AI-assisted programming work.

**Ethical Use of AI:** The responsible application of ChatGPT guided by fairness, transparency, and accountability.

**AI Literacy:** The knowledge and ability to understand, evaluate, and apply AI tools in programming education.

**Institutional Readiness:** The preparedness of an institution to implement and manage AI technologies in education.

**Prompt Engineering:** The design of precise input prompts that guide ChatGPT to produce accurate programming responses.

**Automated Assessment:** An AI-supported process for evaluating and providing feedback on programming tasks.

**Cognitive Dependency:** An overreliance on ChatGPT that weakens independent reasoning and problem-solving effort.

**Human-AI Collaboration:** A cooperative framework where teachers and ChatGPT jointly support learning and evaluation.